\documentclass{article}

\usepackage{hyperref}
\usepackage{tabularx}
\usepackage{array}
\newcolumntype{C}[1]{>{\centering\arraybackslash}m{#1}}
\newcolumntype{L}[1]{>{\raggedright\arraybackslash}m{#1}}
\usepackage{pifont}
\newcommand{\cmark}{\ding{51}}%
\newcommand{\xmark}{\ding{55}}%
\usepackage{microtype}
\usepackage{xcolor}   
\usepackage[nameinlink]{cleveref}
\usepackage[ragged]{footmisc}  

\usepackage{calc}

\usepackage{authblk}

\usepackage{csquotes}
\usepackage[backend=biber, sortcites, defernumbers=true]{biblatex}
\usepackage{graphicx}

\usepackage{breakurl}

\newlength{\itemCol}
\newlength{\descHeight}
\title{SMECS: A Software Metadata Extraction and Curation Software}
\author[1,2]{Stephan Ferenz\thanks{Corresponding Author: \href{mailto:stephan.ferenz@uol.de}{stephan.ferenz@uol.de}}}
\author[2]{Aida Jafarbigloo}
\author[2]{Oliver Werth}
\author[1,2]{Astrid Nieße}
\affil[1]{Department for Computer Science, Carl von Ossietzky Universität Oldenburg, Germany}
\affil[2]{Energy Division, OFFIS, Germany}

\begin{document}
\maketitle

\begin{abstract}
	Metadata play a crucial role in adopting the FAIR principles for research software and enables findability and reusability. However, creating high-quality metadata can be resource-intensive for researchers and research software engineers. 
	To address this challenge, we developed the Software Metadata Extraction and Curation Software (SMECS) which integrates the extraction of metadata from existing sources together with a user-friendly interface for metadata curation. 
	SMECS extracts metadata from online repositories such as GitHub and presents it to researchers through an interactive interface for further curation and export as a \textit{CodeMeta} file. The usability of SMECS was evaluated through usability experiments which confirmed that SMECS provides a satisfactory user experience. SMECS supports the FAIRification of research software by simplifying metadata creation.
\end{abstract}

\section{Introduction}
\label{sec:intro}
Research software plays a vital role in science \cite{gruenpeter_Defining_2021} enabling researchers to simulate real-world systems, control and monitor experiments, prove concepts, and collect, analyze, and interpret complex data \cite{hasselbring_MultiDimensional_2025}.
However, considerable challenges persist when dealing with research software \cite{chuehong_FAIR_2022}. The FAIR (findability, accessibility, interoperability, and reusability) principles for research software \cite{chuehong_FAIR_2022} provide important guidance on how to improve the management of research software and make it FAIR.

An important aspect to achieve FAIR research software is good metadata describing the research software and providing relevant information like contextual knowledge that enables others to understand the purpose, functionality, limitations, and reusability constraints of a research software~\cite{chuehong_FAIR_2022}. 

However, creating meaningful metadata for software is a labor-intensive task as it involves collecting and mostly manually re-entering multiple pieces of information. Although, much of this information is already available but scattered across multiple sources, such as GitHub metadata
or readme files \cite{mao_SoMEF_2019}. Especially when a research software has existed for a long time, the information is highly scattered and often not fully known by a single developer.
To address these challenges, there is a need for tools that can assist researchers and research software engineers (together referred to as users in the following) in creating metadata~\cite{ferenz_Improved_2023}, e.g., by using the available information from the different sources.

Several tools already address different aspects of research software metadata, such as creation (CodeMeta Generator \cite{Codemetagenerator_2025}), extraction (SOMEF \cite{kelley_Framework_2021}), and publication (HERMES~\cite{kernchen_Extending_2025}). However, these existing tools focus on specific aspects and do not provide a comprehensive solution that connects 
all aspects. 

To fill this gap, we present SMECS (Software Metadata Extraction and Curation Software), a novel tool that addresses the challenges of metadata creation for research software. SMECS combines extraction capabilities with a curation interface, aiming for a simple metadata creation process. We evaluated SMECS through a comprehensive user study based on the System Usability Scale (SUS) by Brooke \cite{brooke_sus_1995} and semi-structured interviews.

The remaining part of this paper is organized as follows. In \autoref{sec:relWork}, we provide a comprehensive summary of the related work in the field of research software metadata creation, extraction, and curation, outlining existing tools and their limitations. Afterwards, we introduce our solution, SMECS, and explain its architecture in \autoref{sec:desc}. Based on the evaluation method, described in \autoref{sec:evalMethod}, the results of our evaluation are presented in \autoref{sec:res}. Finally, we discuss and reflect on the implications of our findings and provide an outlook on future research perspectives in \autoref{sec:dis}.

{
\section{Related Work} 
\label{sec:relWork}
This section provides an overview of existing schemas and tools that support the work with research software metadata. We begin by introducing relevant metadata schemas for research software in \autoref{subsec:schemas}, followed by a review of the existing tools in \autoref{subsec:tools}. 

\subsection{Metadata Schemas for Research Software}
\label{subsec:schemas}
There exist multiple metadata schemas for research software. Some of them are domain-specific (e.g., \textit{biotoolsXSD} \cite{ison_BiotoolsSchema_2021}, \textit{Software Ontology} \cite{malone_Software_2014}, \textit{ontosoft} \cite{gil_OntoSoft_2016}) while other are non-domain specific (e.g., \textit{CodeMeta} \cite{jones_CodeMeta_2023}, \textit{CFF} \cite{druskat_Citation_2021}, \textit{Software Description Ontology}~\cite{garijo_OKGSoft_2019}, \textit{DataDesc} \cite{kuckertz_DataDesc_2024}). Ferenz et al. provide a general overview of many of them in~\cite{ferenz_Requirements_2025}. In the following, we will only highlight the most relevant.

The citation file format (\textit{CFF}) \cite{druskat_Citation_2021} provides the most important information for properly citing research software. In this way, it covers some relevant metadata information on research software without defining a proper metadata schema.

\textit{CodeMeta} \cite{jones_CodeMeta_2023} is the de-facto standard for research software metadata covering multiple aspects of research software. It reuses many elements of \textit{schema.org}, which is an ontology describing general concepts and is mainly used to describe things semantically on the web.

The \textit{Software Description Ontology} \cite{garijo_OKGSoft_2019} is an ontology to describe research software, which is based on the geoscience-specific ontology \textit{OntoSoft} \cite{gil_OntoSoft_2016} and \textit{CodeMeta}. Thereby, it provides more metadata elements than \textit{CodeMeta} and better semantic web functionalities.

\subsection{Tools for Curation and Extraction of Research Software Metadata}
\label{subsec:tools}

For the existing tools, we mainly focus on the abilities to extract metadata, allow curation (incl. supporting controlled value vocabularies), and upload the metadata to relevant databases. We do not cover tools for software registries which only focus on storing research software metadata. 

The CodeMeta Generator \cite{Codemetagenerator_2025}, developed by the Software Heritage initiative\footnote{\url{https://www.softwareheritage.org/}, last access 2025-04-07}, is a specialized web tool to simplify the creation of \textit{CodeMeta} metadata, which is available online.\footnote{\url{https://codemeta.github.io/codemeta-generator/}, lass accessed 2025-04-16} The user-friendly form covers most elements of \textit{CodeMeta} and includes drop-down menus for license and development status. The tool also allows users to export a \textit{CodeMeta} file, as well as import and validate existing \textit{CodeMeta} files. The tool does not extract any metadata from any source.

Betty's research engine \cite{seibert_Bettys_2024,seibert_Betty_2024} is a specialized search engine designed to facilitate the discovery of research software. It aggregates structured metadata from multiple sources, including GitHub, GitLab, Zenodo, Open Alex, DataCite, and Open Citations. The engine uses a cascading search approach to iteratively improve the search results. Although, it does not systematically incorporate metadata standards such as \textit{CodeMeta} or \textit{CFF}, it provides a unique platform for searching for research software. Betty's research engine allows users to export search results to the Open Research Knowledge Graph (ORKG) \cite{auer_Improving_2020} with the option to edit the JSON file online before submission \cite{seibert_Bettys_2024}. A demonstrator is available online.\footnote{\url{https://nfdi4ing.rz-housing.tu-clausthal.de/}, last access 2025-04-16}

\begin{table}
\setlength{\extrarowheight}{0.1cm}
	\centering
	\caption{Overview of existing approach to create research software metadata}
        \begin{tabular}{c l c c l}
		\cmark:      & fulfilled         &  & \xmark:     & not fulfilled     \\
		(\cmark):    & (partly) fulfilled  &  &     &    \\
	\end{tabular}

        \begin{tabular}{L{3.8cm} C{2.2cm} C{\itemCol} C{\itemCol} C{\itemCol} C{\itemCol} C{\itemCol}}
		 & 
		Scope & 
		\rotatebox[origin=l]{90}{\parbox[l]{\descHeight}{Extracts structured \\metadata}} & 
        \rotatebox[origin=l]{90}{\parbox[l]{\descHeight}{Extracts unstructured metadata}} & 
        \rotatebox[origin=l]{90}{\parbox[l]{\descHeight}{Curation}} & 
        \rotatebox[origin=l]{90}{\parbox[l]{\descHeight}{Supports value \\vocabularies}} & 
        \rotatebox[origin=l]{90}{\parbox[l]{\descHeight}{Upload of metadata}} \\
		\hline
		\hline
		CodeMeta Generator \cite{Codemetagenerator_2025} & Metadata creation & \xmark & \xmark & \cmark & (\cmark) & \xmark \\
        
        Betty's research engine \cite{seibert_Bettys_2024, seibert_Betty_2024} & Metadata extraction for search & \cmark & \xmark & (\cmark) & \xmark & \cmark \\       
        
		HERMES \cite{druskat_Software_2022, kernchen_Extending_2025, meinel_Hermes_2025} & Metadata publication & \cmark & \xmark & (\cmark) & \xmark & \cmark \\
		
		SOMEF \cite{kelley_Framework_2021, mao_SoMEF_2019} & Metadata extraction & \cmark & \cmark & \xmark & \xmark & \xmark \\
		
	\end{tabular}
	\centering
	\label{Tab:rw}
\end{table}

HERMES \cite{druskat_Software_2022,kernchen_Extending_2025,meinel_Hermes_2025} is a tool for automating the publication of metadata using Continuous Integration/Continuous Deployment (CI/CD) pipelines. Its workflow consists of five key steps: harvest, process, curate, deposit, and post-process. HERMES uses \textit{CodeMeta} as a data model to store metadata between the steps. In the harvest step, it extracts structured metadata from \textit{CFF} files, \textit{CodeMeta} files, local git repositories, and pyproject.toml files. The curation step is facilitated through Pull/Merge Requests on GitHub or GitLab, without providing a dedicated curation interface. HERMES allows metadata to be uploaded to InvenioRDM\footnote{\url{https://inveniosoftware.org/products/rdm/}, last access 2025-04-16} based platforms such as Zenodo \cite{zenodo}. Overall, HERMES is built in a modular way by allowing plugins to be added for all steps of the workflow. This enables easy extendability \cite{kernchen_Extending_2025} and longer interoperability with these extensions.

SOMEF \cite{mao_SoMEF_2019,kelley_Framework_2021,garijo_Framework_2025} is a tool designed to capture metadata from software projects and export it in a structured manner. Its primary goal is to provide broad extraction mechanisms that extract both structured and unstructured metadata. The sources for the structured metadata include the GitHub API and various package management files such as setup.py, pyproject.toml, and package.json. Unstructured metadata are leveraged from README files, licenses, and Docker files. SOMEF does not provide any curation capabilities. It can export metadata as a \textit{CodeMeta} file (JSON-LD) and as an RDF file using the \textit{Software Description Ontology}. The tool is designed to achieve high semantic-web compatibility, enabling the creation of knowledge graphs, as demonstrated in \cite{kelley_Framework_2021}. There is also a web demonstrator for SOMEF called SOMEF-Vider \cite{victor_SOMEF_2021} which is available online\footnote{\url{https://somef.linkeddata.es/}, last access 2025-04-07} \cite{kelley_Framework_2021}.

In this section, we provided an overview of existing tools supporting the creation, extraction, and/or curation of research software metadata as summarized in \autoref{Tab:rw}. While the CodeMeta Generator and SOMEF focus on specific aspects of the process, namely metadata creation and extraction, HERMES covers the entire workflow, including extracting, curating, and publication. However, HERMES lacks a user-friendly curation interface and has limited extraction capabilities compared to SOMEF. Our goal is to bridge this gap by developing a comprehensive tool that spans the entire process, features an intuitive interface, and reuses the existing tools as far as possible. Thereby, we would like to address the current limitations and provide a more seamless experience for users. 
}
{

\section{Description of SMECS}
\label{sec:desc}

We developed SMECS to combine the extraction and curation of research software metadata. Therefore, the workflow of SMECS consists of four sequential phases: \textbf{start}, \textbf{extraction}, \textbf{curation}, and \textbf{export}, as illustrated in \autoref{fig:diagram_smecs_colored}. 
Generally, SMECS addresses the individuals developing research software as main user group. They may be researchers and/or research software engineers. 

In the \textbf{start phase}, users provide two initial inputs. The associated initial interface is shown in \autoref{fig:SMECS_Initial_Page}. The first input prompts users to enter the repository link from 
GitHub, from which metadata extraction is desired. The second input requests the user’s personal token key from the corresponding platform. The tool can also operate without user-provided tokens for certain repositories, using internal default tokens when needed.
    \begin{figure}[t]      
        \includegraphics[width=\textwidth, trim=20 180 50 80, clip]{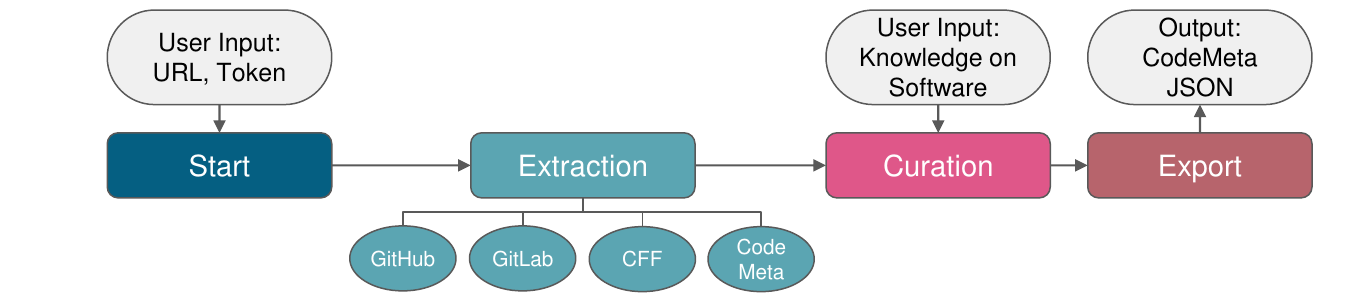}
        \caption{Phases of SMECS}
        \label{fig:diagram_smecs_colored} 
    \end{figure}
    
Based on the entered input, metadata from the repositories are extracted in the \textbf{extraction phase}, which is further outlined in \autoref{subsec:extraction}. Afterwards, the extracted metadata are shown to the user in the \textbf{curation phase}. In this phase, the user can curate and refine the extracted metadata and add additional metadata. We provide more details on this phase and the user interface in \autoref{subsec:curation}. 

Once completed, the curated metadata can be downloaded as a \textit{CodeMeta}-based JSON file in the \textbf{export phase}. The exported metadata can be included by the user in their repository or be used for other purposes.

SMECS is available on GitHub \cite{ferenz_Software_2025} under an AGPL license. The GitHub repository also includes the setup instructions. Additionally, the latest version is available on Zenodo \cite{ferenz_Software_2025a}. 

SMECS is based on the Meta Tool by Reiner Lemoine Institut \cite{RLI_metatool}, which is a web application enabling the creation and validation of metadata for research data based on the Open Energy Metadata standard. Building on this foundation, we developed SMECS as a Python-based tool using the Django framework \cite{Django_2023}. The web frontend uses JavaScript, HTML, and CSS.

     \begin{figure}[h]
        \centering
        \includegraphics[width=\textwidth]{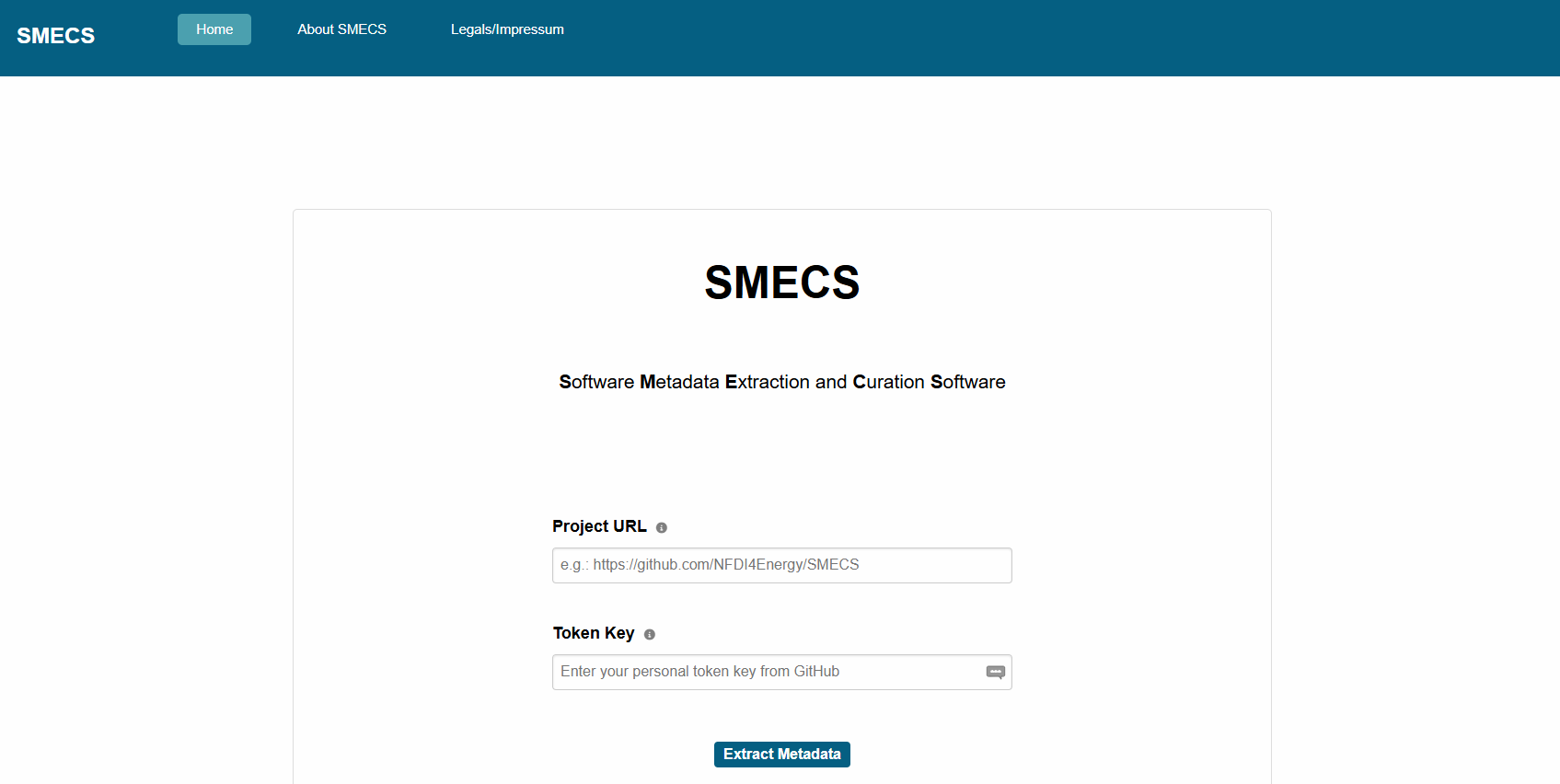}
        \caption{SMECS landing page for the \textbf{start phase}}
        \label{fig:SMECS_Initial_Page}
    \end{figure}

\subsection{Extraction Phase}

\label{subsec:extraction}

Upon entering the input data on the initial page in the \textbf{start phase} and initiating the \textbf{extraction phase}, the tool retrieves metadata. This process uses parts of HERMES \cite{druskat_Software_2022,kernchen_Extending_2025,meinel_Hermes_2025} which serves as the core of the backend as shown in \autoref{fig:extraction-via-hermes}. 
Only small changes to HERMES were performed which will be included in the official HERMES release. For additional functionalities, plugins for HERMES were created which allow long-term interoperability with future HERMES versions.

  \begin{figure}[t]
        \includegraphics[width=\textwidth, trim=20 65 50 75, clip]{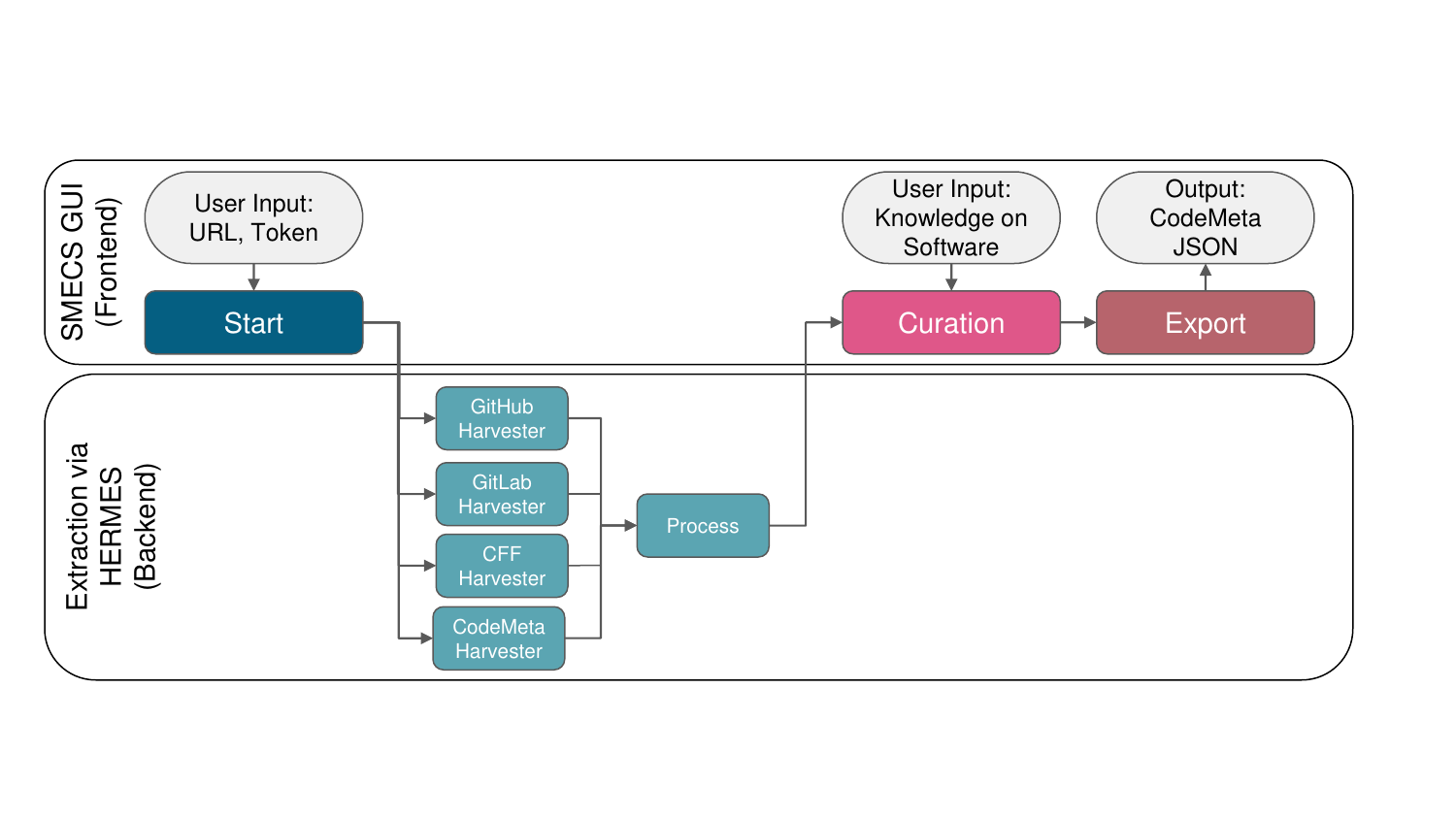}
        \caption{Process diagram: Extracting metadata through using HERMES}
        \label{fig:extraction-via-hermes}
  \end{figure}

  \begin{table}[b]
    \centering
\caption{Supported extraction sources via HERMES}
\label{tab:extraction-sources-via-hermes}
    \begin{tabular}{C{1.4cm} C{4.5cm} C{4cm}} 
         Source& Description & Source for mapping\\ \hline
         GitHub& Metadata are extracted from GitHub API & Official \textit{CodeMeta} crosswalk\footref{fn:cross}\\ 
         \textit{CFF}& \textit{CFF} files in a repository & Crosswalk by HERMES\\ 
         \textit{CodeMeta}& \textit{CodeMeta} files in a repository &  No Crosswalk needed\\ 
    \end{tabular}
\end{table}   

First, metadata are extracted from different sources using the harvesting step of HERMES. The extraction process is done by using three harvesters in HERMES (GitHub, 
\textit{CFF}, and \textit{CodeMeta}). \autoref{tab:extraction-sources-via-hermes} gives an overview of the used sources for metadata. We extended the existing \textit{CFF} and \textit{CodeMeta} harvesters of HERMES to deal with remote repositories and implemented a new harvester for GitHub.
Metadata extracted from GitHub API are more focused on project information (such as repository name, description, and programming languages), collaboration, and activity (such as issues, pull requests, commits, and contributors), however, these sources do not cover citation metadata. On the other hand, \textit{CFF} and \textit{CodeMeta} files are more oriented toward software citation metadata and metadata related to software dissemination and reuse (such as author, versioning, publishers, licensing, and project identifiers like Digital Object Identifier (DOI)). 
All harvesters map their extracted metadata to \textit{CodeMeta}. For this task, we partly used existing crosswalks from \textit{CodeMeta}\footnote{\label{fn:cross}\url{https://codemeta.github.io/crosswalk/github/}, last access 2025-04-10} and HERMES and partly created one crosswalk by ourselves, as shown in \autoref{tab:extraction-sources-via-hermes}. 

The extracted metadata are processed and merged using the process step of HERMES, resulting in a common set of metadata. These metadata are subsequently presented in the user interface in the \textbf{curation phase}. By using HERMES, we achieved an interoperable and modular architecture for the \textbf{extraction phase}, allowing it to easily integrate more harvesting sources.

\subsection{Curation Phase}
\label{subsec:curation}
Based on the extracted metadata from the \textbf{extraction phase}, the extracted metadata are visualized in the \textbf{curation phase}. In \autoref{fig:SMECS_Extraction_Page_1}, this visualization based on an example GitHub URL is showcased. The extracted metadata are organized within an interface resembling a fillable form and are displayed across three tabs: Software Information, Contributors, and Authors. 

Metadata elements, where the automatic extraction was not possible for the provided repository, are visually highlighted in red, indicating where manual input is needed to complete the metadata. Certain metadata elements, like URLs and keyword inputs, are highlighted in yellow when their values are extracted and visualized. This color coding indicates that these elements are suggested to be reviewed or curated by the researcher for more accuracy since the quality of these metadata can often be improved by the users.
 \begin{figure}[h]
        \includegraphics[width=\textwidth]{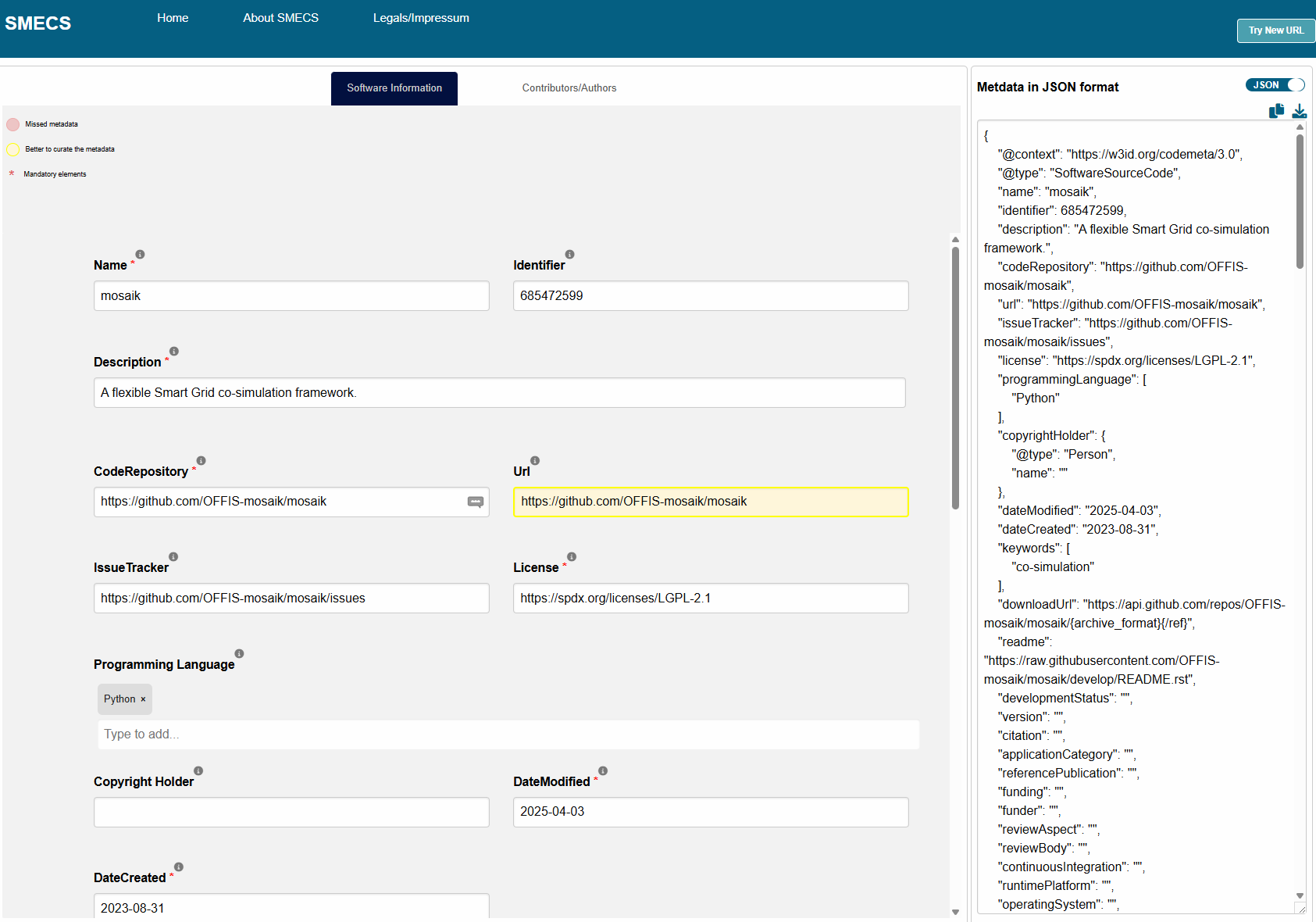}
        \caption{SMECS curation page}
        \label{fig:SMECS_Extraction_Page_1}
  \end{figure}

To enhance the user experience when curating metadata elements such as programming languages and licenses, a list of suggestions is shown to the user. On the frontend, a programming-language JSON file\footnote{\url{https://gist.github.com/calvinfroedge/defeb8fc6cdc0068e172}, last access 2025-04-16} was utilized, offering a comprehensive selection of programming languages. Additionally, a list of licenses\footnote{\url{https://raw.githubusercontent.com/spdx/license-list-data/master/json/licenses.json}, last access 2025-04-16} was fetched to provide users with a valid and up-to-date set of license options for repositories. Dynamic filtering enables users to quickly locate and select an option from a manageable subset, avoiding the overwhelm of a large, unfiltered list.

Once metadata curation is completed in the \textbf{curation phase}, users can download the finalized result as a JSON file or copy that through the copy-to-clipboard feature in the JSON viewer section in the \textbf{export phase}. The \textit{CodeMeta}-based metadata facilitate easy integration with a variety of tools and applications for further use or analysis.

}
{
\section{Research Design and Method, Data Collection and Analysis}
\label{sec:evalMethod}
For the evaluation, we focused on the aspect of usability, a core requirement for SMECS. Accordingly, we applied usability testing that mainly concentrated on the user interface in the \textbf{curation phase}.\footnote{The evaluation was conducted on a previous version of SMECS (Commit number: \href{https://github.com/NFDI4Energy/SMECS/tree/aefc7a92e1a0d092bf51450f637a62a366b7eebd}{aefc7a9}). The main change since the conducted evaluation is the integrating of HERMES which did not influence the frontend.}

Usability testing is a commonly employed method to evaluate how easy a product is to use. This method allows for the identification of problems in the software development stage, focusing on users' ability to understand and interact with the designed functions \cite{guarascio-howard_Using_2004}. Our usability testing aimed to assess how well users could navigate the user interface, understand the tool’s functionalities, and interact with the available features. 

Our usability testing was structured in the following way. First, we gave the participants a short introduction to the session including promising confidentiality of the interview transcripts to avoid possible response biases \cite{myers_Qualitative_2007}. Afterwards, the participants were asked to perform predefined tasks designed to evaluate key aspects of the user experience while also covering the main features of the tool. In this phase, the think-aloud method was used to capture feedback, insights, and thoughts. In the think-aloud method, participants verbalize their thoughts while completing a specific task or recall their thoughts immediately after finishing the task~\cite{eccles_Think_2017}.
All the ideas and thoughts expressed during the think-aloud method were audio-recorded in the session while testing the tool. This qualitative data was analyzed and documented in the same way as the data collected through semi-structured interviews.

To evaluate participants' perception of the user interface of SMECS, quantitative data was gathered through the System Usability Scale (SUS) questionnaire by Brooke \cite{brooke_sus_1995}. Employing the SUS allows for an evaluation of the perceived usability of a system with a limited sample size, providing a reliable assessment of user perceptions of a system or product \cite{tullis_Comparison_2004}. The SUS questionnaire consists of ten statements, each on a five-point scale from ``Strongly Disagree'' to ``Strongly Agree''~\cite{bangor_Determining_2009}, and is used to assess usability in websites, software, and mobile applications~\cite{peres_Validation_2013}.

Additional qualitative data on the tool's usability, functionality, and overall effectiveness was obtained through semi-structured interviews with participants after they completed the SUS questionnaire. The interview guide covered participants’ backgrounds, initial expectations, first-time experiences, specific features, encountered challenges, and overall impressions. 

We recruited all participants from our own research institute\footnote{OFFIS - Institute for Information Technology, \url{https://www.offis.de/}, last access 2025-04-16} to conduct the study on-site, which enabled us to let the participants directly use SMECS on a test machine. All six participants (P1 to P6) were energy researchers at PhD level and engaged in medium to large-scale energy-related research projects. 

They mostly held master's degrees in computer science, with one participant holding a degree in engineering physics. The participants had varying levels of familiarity with metadata, rating their knowledge between 2 and 4 out of 5 (5 = most knowledge). 

All experiments were conducted in English between September and October 2024 using the experimenter's laptop in an on-site setting. Each session lasted between 45 and 60 minutes.
The audios of the interviews were recorded and later transcribed using Whisper \cite{radford_Robust_2022}.
After transcribing the audio recordings, we reviewed the transcripts, highlighted key points, and categorized the feedback into positive and negative opinions, reported bugs, and participant suggestions.

}

{
\section{Results}
\label{sec:res}
Based on the evaluation method explained in \autoref{sec:evalMethod}, the main results of our study are presented in this section.

The SUS scores from the usability experiment, illustrated in \autoref{fig:sus_score_boxplot} as box plot\footnote{\label{fn:plot}The box plot was generated using \cite{blattgerste_WebBased_2022}.}, indicate that participants had a very positive experience using the tool. The scores are consistently high, with a mean of 92.08 and a median of 92.5, reflecting strong user satisfaction. Five participants rated the tool within the ``Excellent'' range (above 85, based on \cite{bangor_Determining_2009}), and one participant scored slightly lower, though still indicating high usability. This consistent pattern suggests the tool strongly aligns with user needs and expectations.

\begin{figure}[t]
    \centering
    \includegraphics[width=0.5\textwidth]{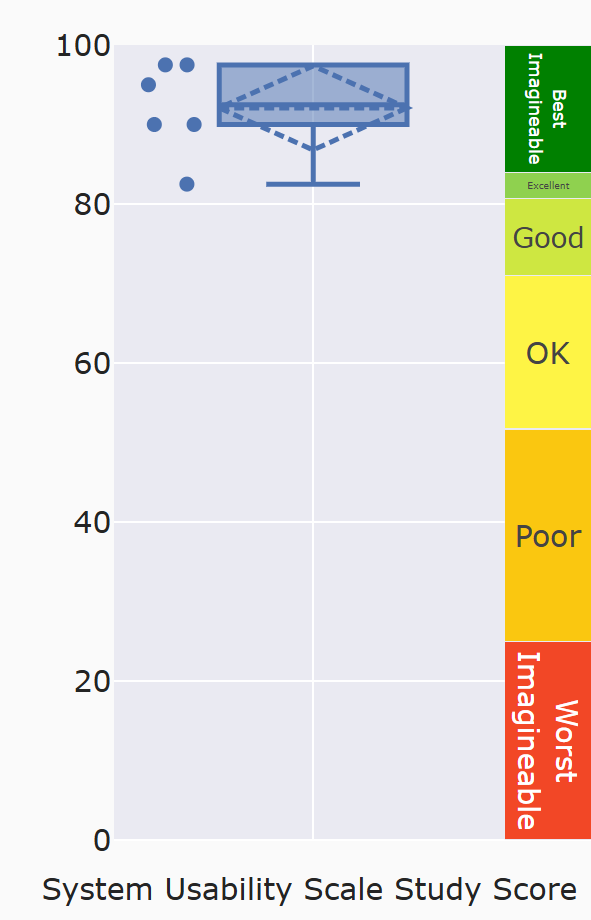}
    \caption{SUS scores for SMECS usability experiment as box plot}
    \label{fig:sus_score_boxplot}
\end{figure}

The qualitative feedback from the think-aloud method and semi-structured interviews offers helpful insights into user experiences and perceptions. Overall, impressions were highly positive with notable comments about clarity, navigation, and ease of use. Participants described the navigation through SMECS as intuitive (P1: ``It's very easy to use, very fast. You can change your information quite fast and it already gives you a lot of information.'', P2: ``The tool is plain, but in a good sense.''). Two participants mentioned that the tool's smooth and responsive functionality mirrors common online tools (P4: ``The tool feels organized, easy to use and feels like an online tool''). 

All participants noted that this segmentation of metadata and visualizing in different categories (tabs) made the page cleaner, organized, and easier to curate relevant metadata (P2: ``The curation form looks really good. It looks clean in my opinion and less intimidating for users that are not technically inclined.'', P5: ``I think it helps not to have too much information on one page, especially when the list of contributors gets very long. If all the metadata is on a single page, you might have to scroll a lot before reaching the next point.''), and they acknowledged the color-coded clues (P3: ``Colors in color coding perfectly matching to the expectation and they are perfectly matching to the standard, the typical things you would see on the Internet and you are used to.''). Participants liked the clarification of roles for contributors and authors, noting that they could easily distinguish these roles when adding or curating the corresponding metadata. All participants appreciated having both options for visualizing metadata during the \textbf{curation phase}, through the form and the JSON viewer (P1: ``The live view of the JSON is pretty nice, and makes it easy to edit metadata in the form and see the result simultaneously.'').

The feedback included suggestions for incorporating advanced features, such as automated license selection and software description generation. Another proposed improvement during the usability experiment was the feature to import a metadata set, obtained from SMECS or other software, into SMECS for the curation of existing research software metadata. Additionally, other potential improvements include implementing validation for specific metadata types and verifying the metadata presented in the JSON viewer. Furthermore, there were recommendations on including detailed information about authors and contributors, specifying what each provided. This can include whether they worked on code, documentation, writing, etc., similar to the way academic publications clarify the roles of each author. 

As part of further development on the features and functionalities, we removed the copy feature from the contributors table. This feature was intended to help users to add a contributor to the authors table, however, its purpose was mostly unclear and misunderstood in the experiment. Most users did not understand the main purpose of this feature. Because of that, we introduced a different approach and decided to merge contributors and authors, presenting related information in a single table for both. This new design allows users to select or deselect roles, meaning a person can be defined as a contributor, an author, or both.
}

{
\section{Discussion and Outlook} 
\label{sec:dis}

Although SMECS has already been rated as usable by researchers in the evaluation (see \autoref{sec:res}), the evaluation and its comparison to related work (see \autoref{sec:relWork}) showed several areas where it can be further extended.

Firstly, multiple existing functionalities can be further expanded. The extraction functionality can be enhanced, for example, by integrating SOMEF. Furthermore, the export functionality, currently limited to downloading a JSON file, can be extended, e.g., to export to the ORKG (similar to Betty's research engine), software repositories (e.g., GitHub and GitLab, as merge requests like in HERMES), and/or software registries (similar to HERMES' upload to invenioRDM). 

Secondly, there are also additional use cases to use SMECS. Besides the export to software registries, it would also be interesting to test SMECS as an input tool for registries by directly integrating it into the registries as envisioned in \cite{ferenz_Improved_2023}.
Furthermore, software management plans are an emerging topic in the area of research software~\cite{alves_ELIXIR_2021}. Since software management plans contain a lot of metadata about research software, it would be an interesting approach to use SMECS or parts of it to semi-automate their creation. 

Additionally, there is potential for utilizing SMECS for more specific metadata schemas, such as the one currently being developed for energy research software \cite{ferenz_Improved_2023, ferenz_Requirements_2025}. 

Our user study as presented in \autoref{sec:evalMethod} and \autoref{sec:res} was designed to evaluate the usability and effectiveness of SMECS. While the results provide valuable insights into the performance of the tool, some limitations of the evaluation must be acknowledged. 

Firstly, our study had a small number of participants with homogeneous backgrounds in computer science, which may have introduced biases in the results. Despite this limitation, we were able to gain a good understanding of the tool's usability and have already incorporated some of the findings (e.g., metadata from existing \textit{CodeMeta} files in a repository can now be extracted). Future studies should aim to recruit a more diverse sample to ensure that the results are generalizable to a broader group of researchers.

In addition, the use of predefined tasks may have limited the scope of the evaluation, but it also allowed us to test the usability of specific features of SMECS. 
Furthermore, the most recent version of the tool was not used in the usability study. Consequently, the findings may not be entirely applicable to this latest version.

Overall, further research is necessary to explore the capabilities of SMECS and its usability by researchers. 

Also, the general impact of tools like SMECS on researchers' behavior with respect to the metadata creation remains an area for further exploration. Understanding the effectiveness of SMECS and similar tools in promoting metadata creation and publication is relevant for developing strategies to increase the adoption of metadata for research software.  

In conclusion, we showed that SMECS is a usable tool for creating metadata for research software based on extraction. By building on HERMES, it achieves a high degree of interoperability with existing tools. Nevertheless, there are still opportunities for improvement, particularly in refining its extraction and export capabilities. Additionally, more research, such as integrating other metadata schemas or conducting studies with more diverse participants, is needed to better understand what is needed to enable researchers to create meaningful metadata for their research software.
}

\section*{Acknowledgments}
The authors would like to thank the German Federal Government, the German State Governments, and the Joint Science Conference (GWK) for their funding and support as part of the \href{https://nfdi4energy.uol.de/}{NFDI4Energy} and the \href{https://base4nfdi.de/}{Base4NFDI} consortia. The work was partly funded by the German Research Foundation (DFG) – 501865131 and 521453681 within the German National Research Data Infrastructure (NFDI, \url{www.nfdi.de}).

Additionally, this research was partly funded by the Lower Saxony Ministry of Science and Culture under grant number 11-76251-13-3/19–ZN3488 (ZLE) within the Lower Saxony “Vorab“ of the Volkswagen Foundation. It was supported by the \href{https://zdin.de/}{Center for Digital Innovations Lower Saxony (ZDIN)}.

\printbibliography[title=References]

\end{document}